\documentclass[12pt]{article}
\usepackage{amsmath,amsfonts,latexsym,url}

\newenvironment{proof}{
\par
\noindent {\bf Proof.}\rm}%
{\hspace*{\fill}\rule{0.5em}{0.809em}\par}

\newcommand{\cross}{\otimes}

\newcommand{\ket}[1]{\mbox{$|#1\rangle$}}

\def\ceil#1{\lceil#1\rceil}

\newcommand{\PrS}{{\textstyle \Pr_{\cal S}}}

\newcommand{\squash}[1]{\raisebox{0.04ex}[0pt][0pt]{\small$\textstyle #1$}}
\newcommand{\oosrt}{\squash{\frac{1}{\sqrt{2}}}}

\newcommand{\optprob}{\mbox{\smash{$\frac{1}{2} + 2^{-\ceil{n/2}}$}}}

\newcommand{\pbfrac}[2]{\mbox{$\mbox{}^{#1}\!/_{#2}$}}
\newcommand{\mypmod}[1]{~(\textup{mod}~#1)}

\newcommand{\win}{\textnormal{win}}

\begin{document}

\title{Recasting {M}ermin's multi-player game\\
into the framework of pseudo-telepathy}

\author{\large Gilles Brassard\,%
\thanks{\,Supported in part by Canada's Natural Sciences and Engineering
Research Council (NSERC),
the~Canada Research Chair \mbox{programme} and
the Canadian Institute for Advanced Research (CIAR).}
~~~~~Anne Broadbent\,%
\thanks{\,Supported in part by a scholarship from Canada's NSERC.}
~~~~~Alain Tapp\,%
\thanks{\,Supported in part by Canada's NSERC,
Qu\'ebec's Fonds de recherche sur la nature et les technologies (FQRNT),
the CIAR and the
Mathematics of Information Technology and Complex Systems Network (MITACS).}
\\ {\normalsize\it D\'epartement~IRO, Universit\'e de Montr\'eal}\\[-1ex]
{\normalsize\it C.P.~6128, succursale centre-ville}\\[-1ex]
{\normalsize\it Montr\'eal (Qu\'ebec),  H3C~3J7 \textsc{Canada}}\\
{\normalsize\texttt{\{brassard,\,broadbea,\,tappa\}}\textbf{\char"40}\texttt{iro.umontreal.ca}}}

\date{Revised 14 June 2005}

\newtheorem{theorem}{Theorem} 
\newtheorem{corollary}[theorem]{Corollary}
\newtheorem{lemma}[theorem]{Lemma}
\newtheorem{proposition}[theorem]{Proposition}
\newtheorem{definition}{Definition}
\newtheorem{notation}[definition]{Notation}

\maketitle

\sloppy

\begin{abstract}

Entanglement is perhaps the most non-classical manifestation of
quantum \mbox{mechanics}.  Among its many interesting applications
to information processing, it can
be harnessed to \emph{reduce} the amount of communication
required to process a variety of \mbox{distributed} computational tasks.
Can~it be used to \emph{eliminate} communication altogether?
Even though it cannot serve to signal information between remote parties,
there are distributed tasks that can be performed without any need
for communication, provided the parties share prior entanglement:
this is the realm of \emph{pseudo-telepathy}.

One of the earliest uses of \mbox{multi-party} entanglement was presented
by Mermin in 1990.  Here we recast his idea in terms of pseudo-telepathy:
we~provide a new computer-scientist-friendly analysis of this game.
We~prove an upper bound on the best possible
classical strategy for attempting to play this game, as well as a novel, matching lower bound.
  This leads us
to considerations on how well imperfect quantum-mechanical apparatus
must perform in order to exhibit a behaviour that would be
classically impossible to explain. Our~results include improved bounds that
could help vanquish the infamous detection loophole.

\end{abstract}

\section{Introduction}
\label{introduction}

It is well-known that quantum mechanics can be harnessed to reduce
the amount of communication required to perform a variety of distributed
tasks, by clever use of \mbox{either} quantum communication~\cite{CDNT98}
(in~the model of Yao~\cite{yao83}) or quantum entan\-gle\-ment~\cite{CB}.
Consider for \mbox{example} the case of Alice and Bob, two very busy scientists
who would like to find a time when they are simultaneously free for lunch.
They each have an engagement \mbox{calendar}, which we may think of as
\mbox{$n$-bit} strings $a$ and~$b$, where
\mbox{$a_i=1$} (resp.~\mbox{$b_i=1$}) means that Alice (resp.~Bob) is free for
lunch on day~$i$.  Mathematically, they want to find an index~$i$ such that
\mbox{$a_i=b_i=1$} or establish that such an index does not \mbox{exist}.
The obvious solution is for Alice, say, to communicate her entire calendar
to Bob, so that he can decide on the date: this requires roughly
$n$ bits of communication.  It~turns out that this is optimal in the worst
case, up to a constant factor, according to classical information
\mbox{theory}~\cite{KS}, even when the answer is only required to be correct
with probability at least~\pbfrac{2}{3}.  Yet,~this problem can be solved with
arbitrarily high success probability with the \mbox{exchange} of a number of
\emph{quantum} bits---known as \emph{qubits}---in
the order of~$\sqrt{n}$~\cite{aaronson}.
Alternatively, a number of \emph{classical} bits in the order of
$\sqrt{n}$ \mbox{suffices} for this task if Alice and Bob share prior entanglement,
because they can make use of quantum teleportation~\cite{teleport}.
Other (less natural) problems demonstrate an \emph{exponential}
advantage of quantum communication, both in the error-free~\cite{BCW} and
bounded-error~\cite{raz} models.
Please consult~\cite{survey,deWolf} for surveys on the topic
of quantum communication \mbox{complexity}.
\looseness=+1

Given that prior entanglement allows for a dramatic \emph{reduction} in the
need for \mbox{classical} communication in order to perform some distributed
computational tasks, it is natural to wonder if it can
be used to \emph{eliminate} the need for communication altogether.
In~other words, are there distributed tasks
that would be impossible to achieve in a classical world if the participants were
not allowed to communicate, yet those tasks could be performed without
\emph{any} form of communication provided they share prior
entanglement? The answer is negative if the result of the computation must
\mbox{become} known to at least one party---otherwise, this phenomenon
could be harnessed to provide faster-than-light signalling.
Nevertheless, the feat becomes possible if we are satisfied with the
establishment of nonlocal \emph{correlations} between the parties'
inputs and \mbox{outputs}~\cite{BCT99}.
\looseness=+1

Mathematically, consider $n$ parties $A_1$, $A_2$,\ldots, $A_n$,
called the \emph{players}, and two \mbox{$n$-ary} functions $f$
and~$g$\@. In~an \emph{initialization phase}, the players are
allowed to discuss strategy and share random variables (in the
classical setting) and entanglement (in the quantum setting). Then
the players move apart and they are no longer allowed any form of
communication. After the players are physically separated, each
$A_i$ is given some input $x_i$ and is requested to produce
output~$y_i$. We~say that the players \emph{win} this instance of
the game if
\mbox{$g(y_1,\,y_2,\ldots\,y_n)=f(x_1,\,x_2,\ldots\,x_n)$}. Given
an $n$-ary predicate~$P$\!, known as the \emph{promise}, a
strategy is \emph{perfect} if it wins the game with certainty on
all questions that satisfy the promise, i.e.~whenever
\mbox{$P(x_1,\,x_2,\ldots\,x_n)$} holds. A~strategy is
\emph{successful with probability}~$p$ if it wins \emph{any}
instance that satisfies the promise with probability at least~$p$;
it~is successful in \emph{proportion} $p$ if it wins the game with
probability at least $p$ when the instance is chosen at random
according to the uniform distribution on the set of instances that
satisfy the promise. Any strategy that succeeds with probability
$p$ automatically succeeds in proportion~$p$, but not necessarily
vice versa. In~particular, it is possible for a strategy that
succeeds in proportion $p>0$ to fail systematically on some
questions, whereas this would not be allowed for strategies that
succeed with probability~$p>0$. Therefore, the notion of
succeeding in~proportion is the only one that is meaningful for
\emph{deterministic} strategies, and this is indeed where the name
``in~proportion'' comes from: it is the ratio of the number of
questions on which the strategy provides a correct answer to the
total number of possible questions, taking account only of
questions $x_1 x_2 \cdots x_n$ for which $P(x_1, x_2, \ldots,
x_n)$ holds.

We say of a quantum strategy that it exhibits
\emph{pseudo-telepathy} if it is perfect provided the players
share prior entanglement, whereas no perfect classical strategy
can exist. The~study of pseudo-telepathy was initiated
in~\cite{BCT99}, but games that fit this
framework had been introduced earlier~\cite{HR82,GHZ} (but~\emph{not} \cite{hardy}, see~\cite{BMT04}).
Unfortunately, those earlier papers were presented in a physics jargon
hardly accessible to computer scientists, even with \mbox{decent}
background in quantum information theory.
Mermin offered a refreshing but temporary relief to this
physicists-writing-for-their-kind-only paradigm when he
presented a very accessible three-player account~\cite{MerminGHZ} of the
GHZ scenario~\cite{GHZ}.  This protocol was also set into the communication complexity
framework in~\cite{BCD01}.

But even Mermin donned his physicist's hat when he generalized
his own game to an arbitrary number of players~\cite{Me90} in~1990.
In~this article, we develop the pseudo-telepathy game thus
introduced by Mermin, which involves $n\geq 3$ players.
This is probably the simplest multi-player game possible
because each player is given a single bit of input and is
requested to produce a single bit of output.
Moreover, the quantum perfect strategy requires each player
to handle a single qubit.
To~the best of our knowledge, this 1990 game is also the first
pseudo-telepathy game ever proposed that is \textit{scalable}
to an arbitrary number of players.

We recast Mermin's $n$-player game in terms of pseudo-telepathy in
Section~\ref{quantum} and we give a perfect quantum strategy
for~it. In~Sections~\ref{classical} and~\ref{probabilistic}, we
prove that no classical strategy can succeed with a probability
that differs from random guessing by more than an exponentially
small fraction in the number of players.  More
\mbox{specifically}, no classical strategy can succeed in the
$n$-player game with a probability better than~$\optprob$. Then,
we match this bound with a novel explicit classical strategy that
is successful with the exact same probability~$\optprob$. Finally,
we show in Section~\ref{loophole} that the quantum success
probability would remain better than anything classically
achievable, when $n$ is sufficiently large, even if each player
had imperfect apparatus that would produce the wrong outcome with
probability nearly 15\% or no outcome at all with probability
close to 50\%. This could be used to circumvent the infamous
\emph{detection loophole} in experimental proofs of the
nonlocality of the world in which we live~\cite{massar}. We~assume
throughout this paper that the reader is familiar with elementary
concepts of quantum information processing~\cite{nielsen}.

\section{The Game and its Perfect Quantum Strategy}
\label{quantum}

For any $n \ge 3$, game $G_n$ involves  $n$ players. Each player
$A_i$ receives a single input bit $x_i$ and is requested to
produce a single output bit~$y_i$. The players are promised that
there is an even number of 1s among their inputs. Without being
allowed to communicate after receiving the question, they
are challenged to produce a collective answer that contains an
even number of 1s if and only if the number of 1s in the inputs is
divisible by~4. More formally, we require that
\begin{equation}\label{goal}
\sum_{i=1}^n y_i ~\equiv~
{\textstyle \frac12} {\sum_{i=1}^n x_i} \pmod 2 \,
\end{equation}
provided $\sum_i x_i$ is even.
We say that $x=x_1x_2 \cdots x_n$ is the \emph{question} and
$y=y_1y_2 \cdots y_n$ is
the \emph{answer}, which is \emph{even}
if it contains an even number of 1s and \emph{odd} otherwise.
We say that a question is \emph{legitimate} if it
satisfies the promise and that an answer is \emph{appropriate} if
Equation~\ref{goal} is satisfied.  Please do not confuse the words
``input'' and ``question'': the former refers to the single bit~$x_i$
seen by one of the players whereas the latter refers to the collection~$x$
of all input bits that serves as challenge for the collectivity of
players.  The same distinction applies between
``output'' and ``answer''.

\begin{theorem}  \label{thm:quant}
If the $n$ players are allowed to share prior entanglement, then
they can always win game $G_n$.
\end{theorem}

\begin{proof}
Define the following $n$-qubit entangled quantum states
\ket{\Phi_n^+} and~\ket{\Phi_n^-}:
\begin{align}
\ket{\Phi_n^+} &= \oosrt \ket{0^n} + \oosrt \ket{1^n} \nonumber\\[1.5ex]
\ket{\Phi_n^-} &= \oosrt \ket{0^n} - \oosrt \ket{1^n} \, . \nonumber
\end{align}
Let $H$ denote the Walsh-Hadamard transform, defined as usual by
\[ H ~:~
\begin{cases}
~\ket{0} ~\mapsto~ \oosrt\ket{0}+\oosrt\ket{1} \\[1.5ex]
~\ket{1} ~\mapsto~ \oosrt\ket{0}-\oosrt\ket{1}
\end{cases}
\]
and let $P$ denote a phase-change unitary transformation defined by
\[ P ~:~
\begin{cases}
~\ket{0} ~\mapsto~ \phantom{\imath}\ket{0} \\[1ex]
~\ket{1} ~\mapsto~ \imath \ket{1} \, ,
\end{cases}
\]
where we use a dotless $\imath$ to denote $\sqrt{-1}$ in order to
distinguish it from index $i$, which is used to identify a player.
It~is easy to see that if $P$ is applied to any two qubits of
\ket{\Phi_n^+}, while the other qubits are left undisturbed,
the resulting state is \ket{\Phi_n^-}, and vice versa.
Moreover, if P is applied to any \emph{four} qubits of
\ket{\Phi_n^+} or \ket{\Phi_n^-}, while the other qubits
are left undisturbed, the global state stays the same.
Therefore, if the qubits of \ket{\Phi_n^+} are
distributed among the $n$ players, and if exactly $m$ of them
apply $P$ to their qubit, the resulting global state remains~\ket{\Phi_n^+}
if \mbox{$m \equiv 0 \mypmod4$}, whereas it evolves to~\ket{\Phi_n^-}
if \mbox{$m \equiv 2 \mypmod4$}.

Furthermore, the effect of applying the Walsh-Hadamard transform to each qubit in
\ket{\Phi_n^+} is to produce an equal superposition of all even \mbox{$n$-bit}
strings,
whereas the effect of applying the Walsh-Hadamard transform to each qubit
in \ket{\Phi_n^-} is to produce an equal superposition of all odd \mbox{$n$-bit}
strings.
More formally,
\begin{align}
(H^{\cross n})  \ket{\Phi_n^+} &=
{\textstyle \frac{1}{\sqrt{2^{n-1}}}} \!\!
 \sum_{y\text{~even}} \ket{y} \nonumber \\[2ex]
(H^{\cross n})  \ket{\Phi_n^-} &=
{\textstyle \frac{1}{\sqrt{2^{n-1}}}} \!\!
 \sum_{y\text{~odd}} \ket{y} \nonumber
\end{align}
where $y$ ranges over all \mbox{$n$-bit} strings.

The quantum winning strategy should now be obvious.
In~the initialization phase, the $n$ qubits of state \ket{\Phi_n^+}
are distributed among the $n$ players.
After they have moved apart, each player $A_i$ receives input
bit~$x_i$ and does the following:
\begin{enumerate}
\item\label{stepone} apply transformation $P$ to qubit if \mbox{$x_i = 1$}
(skip this step otherwise);
\item\label{steptwo} apply $H$ to qubit;
\item measure qubit in the computational basis ($\ket0$~versus~$\ket1$)
in order to obtain $y_i$\,;
\item produce $y_i$ as output.
\end{enumerate}

We know by the promise that an even number of players will apply $P$ to
their qubit.  If that number is divisible by~4, which happens when
$\squash{\frac12} {\sum_i x_i}$ is even, then the global
state reverts to \ket{\Phi_n^+} after step~\ref{stepone} and therefore to
a superposition of all \ket{y} such that $y$ is even
after step~\ref{steptwo}.  It~follows that $\sum_i y_i$, the number of players
who measure and output~1, is even. On~the other hand, if
the number of players who apply $P$ to their qubit is congruent to~2
modulo~4, which happens when $\squash{\frac12} {\sum_i x_i}$ is odd, then the
global state evolves to \ket{\Phi_n^-} after step~\ref{stepone} and
therefore to a superposition of all \ket{y} such that $y$
is odd after step~\ref{steptwo}.  It~follows in this case that
$\sum_i y_i$
is odd. In~either case, Equation~\ref{goal} is satisfied at the
end of the strategy, as required. 
\end{proof}

\section{Optimal Proportion for Deterministic Strategies}
\label{classical}

In this section, we prove matching upper and lower bounds on the
success proportion achievable by deterministic strategies that
play game $G_n$ for any~\mbox{$n \ge 3$}.

\begin{theorem} \label{thm:prop}
Any deterministic strategy for game $G_n$ is
successful in proportion at most~$\optprob$.
\end{theorem}

\begin{proof}
Let $S$ be a deterministic strategy specified by $S_{ij}$, where
$S_{ij}=1$ if player $i$'s output on input $j$ is 0 and
$S_{ij}=-1$ otherwise.  Notice that
we can consider the sign of the product of a subset of the
$S_{ij}$s in order to determine if the game is won: for a given
question $x=x_1 x_2 \cdots x_n$, $\prod_{i=1}^n S_{ix_i}=1$ if the
players' answer $y=y_1 y_2 \cdots y_n$ is even and $\prod_{i=1}^n
S_{ix_i}= -1$ if the players' answer is odd.
Consider the following quantity~$s$.
\begin{align}
s  ~=~& \prod_{i=1}^n (S_{i0}+\imath S_{i1}) \label{eqprod} \\
   ~=~& \sum_{x \in \{0,1\}^n}
   \left( \imath^{\Delta(x)} \prod_{i=1}^n S_{i\,x_i}\right) \label{eqsum}
\end{align}
where $\Delta(x) = \sum_i x_i$ denotes the Hamming weight of~$x$
(the~number of 1s in~$x$).
By~\mbox{expanding} the product into a sum, we see that each term
corresponds to an \mbox{$n$-bit} string~$x$.  If $\Delta(x)$ is odd, then
the question $x$ is not legitimate, in which case
$\imath^{\Delta(x)}$ is purely imaginary.  Otherwise, if $x$ is
legitimate, $\imath^{\Delta(x)}$ is real. More to the point,
$\imath^{\Delta(x)} = 1 $ if ${\textstyle \frac12} {\sum_i x_i}$
is even and $\imath^{\Delta(x)} = -1 $ otherwise.

In order for strategy $S$ to give an appropriate answer on
question $x$, we must have that \mbox{$\prod_i S_{i\, x_i} = 1$}
if ${\textstyle \frac12} {\sum_i x_i}$ is even and \mbox{$\prod_i
S_{i\, x_i} = -1$} otherwise.  Combining this with the previous
observations, we conclude that for all legitimate questions, the
corresponding term in the expansion of $s$ (Equation~\ref{eqsum})
is $1$ if the strategy gives an appropriate answer on
question~$x$, and it is $-1$ otherwise. It~follows that
$\text{Re}(s)$, the real part of~$s$, is precisely the number of
appropriate answers minus the number of inappropriate answers
provided by strategy~$S$, counted on the set of all legitimate
questions. To~upper-bound $\text{Re}(s)$, we revert to
Equation~\ref{eqprod}. Consider each factor of the product that
defines~$s$: \mbox{$S_{i0}+\imath S_{i1} = \sqrt{2} e^{\imath a_i
\pi/4}$} for some $a_i$ in \mbox{$\{1,3,5,7\}$}. Thus, if $n$ is
even, we have \mbox{$s \in \{ 2^{n/2},\imath
2^{n/2},-2^{n/2},-\imath 2^{n/2}\}$} and \mbox{$\text{Re}(s) \leq
2^{n/2}$}. If~$n$ is odd, we have \mbox{$s \in \{ 2^{n/2}( \pm
\frac{ 1}{\sqrt{2}} \pm \frac{1}{\sqrt{2}} \imath) \}$} and
\mbox{$\text{Re}(s) = \pm 2^{(n-1)/2}$}. In~either case,
\smash{\mbox{$\text{Re}(s) \le 2^{\left\lfloor n/2
\right\rfloor}$}}.

The difference between the number of appropriate answers and the
number of inappropriate answers is at most
\mbox{$\text{Re}(s) \le 2^{\left\lfloor n/2 \right\rfloor}$},
but the sum of those two numbers is~$2^{n-1}$,
the total number of legitimate questions.
It~follows---by~adding these two statements and dividing by~2---that
the number of appropriate answers is at most
\mbox{$2^{n-2}+2^{\left\lfloor n/2 \right\rfloor-1}$}.
The~desired upper bound on the proportion of appropriate answers is finally
obtained after a division by the number of legitimate questions:
\[
\frac{2^{n-2}+2^{\left\lfloor n/2 \right\rfloor-1}}{2^{n-1}}
~=~ \optprob \, .
\]  
\end{proof}

It turns out that \emph{very} simple deterministic strategies achieve
the bound given in
Theorem~\ref{thm:prop}. In~particular, the players do not even
have to look at their input when \mbox{$n \not\equiv 2 \mypmod 4$}.
Even when \mbox{$n \equiv 2 \mypmod 4$}, it is sufficient for a single
player to look at his input!

\begin{theorem} \label{thm:achievable}
There is a classical deterministic strategy for game $G_n$ that is
successful in proportion exactly~$\optprob$.
\end{theorem}

\begin{proof}
A tedious but straightforward case analysis suffices to establish
that the following simple strategies (Table \ref{table:win}),
which depend on $n \mypmod 8$, succeed in proportion
exactly~$\optprob$. We~have used two bits to represent a player's
strategy, where the first bit of the pair denotes the strategy's
output $y_i$ if the input bit is \mbox{$x_i=0$} and the second bit
of the strategy denotes its output if the input is \mbox{$x_i=1$}.
(For example, player~1 would output \mbox{$y_1=0$} on input
\mbox{$x_1=1$} if $n$ is congruent to~6 modulo~8.) A~pair of
identical bits as strategy means that the corresponding player
outputs that bit regardless of his input bit.
\end{proof}
\begin{table}[ht]
\caption{\label{table:win}Simple optimal strategies.}
\centering
\begin{tabular}{|c|c|c|}\hline
 $n \mypmod 8$ & player 1  & players 2 to $n$    \\ \hline
  0   &  00      &        00       \\ \hline
  1   &  00      &        00       \\ \hline
  2   &  01      &        00       \\ \hline
  3   &  11      &        11       \\ \hline
  4   &  11      &        00       \\ \hline
  5   &  00      &        00       \\ \hline
  6   &  10      &        00       \\ \hline
  7   &  11      &        11       \\ \hline
\end{tabular}
\end{table}

\section{Optimal Probability for Classical Strategies}
\label{probabilistic}

In this section, we consider all possible \emph{classical}
strategies to play game $G_n$, \mbox{including} probabilistic
strategies. We~give as much power as possible to the classical
model by allowing the playing parties unlimited sharing of random
variables. Despite this, we prove that no classical strategy can
succeed with a probability that is significantly better
than~\pbfrac{1}{2} on the worst-case question, and we show that
our lower bound is tight by exhibiting a probabilistic
classical strategy that achieves~it.

\begin{definition}\label{strategy}
A probabilistic strategy $\cal S$ is a probability distribution
over a finite set of deterministic strategies.
\end{definition}

Without loss of generality, the random variables shared by the players
during the initialization phase correspond to deciding which
deterministic strategy will be used for any given instance of the
game.

\begin{notation}
Given an arbitrary strategy $\cal S$ and legitimate question~$x$,
let \mbox{$\PrS(\win \mid x)$} denote the probability
that strategy $\cal S$ provides an appropriate answer on question~$x$,
and let
\[
\PrS(\win) =
\frac{1}{2^{n-1}} \sum_x \PrS(\win \mid x)
\]
denote the average success probability of strategy {\cal S}
when the question is chosen at random according to the uniform
distribution among all legitimate questions.
\end{notation}

Whenever $S$ is a deterministic strategy, note
that \mbox{$\Pr_S(\win \mid x) \in \{0,1\}$}
and \mbox{$\Pr_S(\win)$} is the same as what we had called
the success proportion.
If~$\cal S$ is a probabilistic strategy,
\mbox{$\PrS(\win)$} corresponds also to the success proportion,
which is not to be confused with the more interesting notion of
success \emph{probability}.  Indeed, the formal definition
of the success probability of $\cal S$ involves taking the
\emph{minimum} rather than the average of the $\PrS(\win \mid x)$
over all~$x$.

It is well known~\cite{yao77}
that the success probability of an arbitrary
classical strategy, even probabilistic, can never exceed the success
proportion of the best possible deterministic strategy
(for the case of pseudo-telepathy, this is proved in~\cite{BBT04b}). Even though
Theorem~\ref{framework:probabilistic} (below) follows directly from this
general principle,
we~give it an explicit proof for the sake of completeness.

\begin{theorem}\label{framework:probabilistic}
Any classical strategy for game $G_n$ is successful
with probability at most~$\optprob$.
\end{theorem}
\begin{proof}
Consider a general probabilistic strategy  $\cal S$, which is a
probability distribution over deterministic strategies
\mbox{$\{s_1, s_2, \ldots, s_\ell \}$}. Let~$\Pr(s_j)$ be the
probability that strategy $s_j$ be chosen on any given instance of
the game. Let~$p$ be the success probability of~$\cal S$, which is
the quantity of interest in this theorem. By~definition, \mbox{$ p
\leq \PrS(\win \mid x)$} for any legitimate question~$x$, and
therefore \mbox{$ p \leq \PrS(\win)$} as well. (This~simply says
that the minimum can never exceed the average.) Also by
definition,
\[ \PrS(\win \mid x) = \sum_j \Pr(s_j) \, \Pr_{s_j}(\win \mid x) \, . \]
Putting it all together,
\begin{eqnarray*}
  p &\leq&  \PrS(\win) \\[1ex]
    & = & \frac{1}{2^{n-1}} \sum_{x} \PrS(\win \mid x) \\
  &=& \frac{1}{2^{n-1}} \sum_{x} \sum_j  \Pr(s_j) \, {\textstyle \Pr_{s_j}}( \win \mid x) \\
  &=&  \sum_j   \Pr(s_j)   \frac{1}{2^{n-1}} \sum_{x}  {\textstyle \Pr_{s_j}}( \win \mid x) \\
  &=&  \sum_j   \Pr(s_j) \, {\textstyle \Pr_{s_j}}( \win ) \\
  &\leq& \sum_j \Pr(s_j) \left(  \optprob \right) \\[2ex]
  & = & \optprob \, .
\end{eqnarray*}
The last inequality comes from Theorem~\ref{thm:prop}.
\end{proof}

We now proceed to prove that Theorem~\ref{framework:probabilistic}
is tight.

\begin{definition}We define an \emph{optimal strategy} to be a deterministic
strategy that is successful in proportion exactly~$\optprob$.
\end{definition}

We know from Theorem~\ref{thm:achievable} that optimal strategies
exist and from Theorem~\ref{thm:prop} that they are optimal indeed.

\begin{definition}
A set $O$ of optimal strategies is \emph{balanced} if the number
of strategies in $O$ that answer appropriately any given
legitimate question is the same for each legitimate question.
\end{definition}

Note that it is not \emph{a~priori} obvious that nontrivial
balanced sets of optimal strategies exist at~all.  We~shall prove this
later, but let us take them for granted for now.

\begin{lemma}\label{lem:probabilistic}
Consider any nonempty balanced set $O$ of optimal strategies.
Define probabilistic strategy
$\mathcal{S}$ for game $G_n$ as a uniform distribution
over~$O$. Then $\mathcal{S}$
is successful with probability $\optprob$.
\end{lemma}

\begin{proof}
Consider the proof of Theorem~\ref{framework:probabilistic}.
Because $O$ is balanced, $\PrS(\win \mid x)$ is the same
for all~$x$, and therefore the average of these values is
the same as their minimum.  This means that if $p$ is the
success probability of $\cal S$, then \mbox{$p = \PrS(\win)$} as well.
Moreover, \mbox{$\Pr_{s_j}(\win) = \optprob$} for each~$j$ because
each $s_j$ is optimal.  It~follows that
both inequalities in the proof of Theorem~\ref{framework:probabilistic}
become equalities, and therefore the success
probability of $\mathcal{S}$ is $\optprob$. 
\end{proof}

\begin{theorem}\label{thm:acheaveprob}
There is a classical probabilistic strategy for game $G_n$ that is
successful with probability exactly $\optprob$.
\end{theorem}

\begin{proof}
Consider the probabilistic strategy $\mathcal{S}$ that is a
uniform distribution over the set $O$ of \emph{all} optimal
strategies.  If we show that $O$ is balanced,
then it follows by Lemma \ref{lem:probabilistic} that
$\mathcal{S}$ is successful with probability~$\optprob$.

Using the
same notation as in Theorem~\ref{thm:prop}, a deterministic strategy $S$
is optimal if and only if
\[  \text{Re} \left[ \prod_{i=1}^n (S_{i0}+\imath S_{i1}) \right] =
 2^{\left\lfloor n/2 \right\rfloor} \, .
\]
We proceed to show that if we flip two bits of any legitimate question, we get
another legitimate question for which there are at least as many optimal
strategies that give an appropriate answer.
Because it is possible to go from any legitimate question to any
other legitimate question by a sequence of two-bit flips,
this shows that the number of optimal strategies that give an
appropriate answer is the same for all legitimate questions.

Assume without lost of generality that the two questions differ in
the first two positions.  Assume furthermore that $x= 0 0 x_3
\cdots x_n$ and $x'=1 1 x_3 \cdots x_n$. (A~similar reasoning
works if the first two bits of $x$ are $01,10$ or~$11$, or if the
two questions differ in any other two positions.)

To each optimal strategy $S$ that gives an appropriate answer
for~$x$, we associate a strategy $S'$ that gives an appropriate
answer for~$x'$. The mapping that does the association between the
strategies is a one-to-one correspondence defined as follows:
\mbox{$S'_{10}=S_{11}$}, \mbox{$S'_{11}=-S_{10}$},
\mbox{$S'_{20}=-S_{21}$}, \mbox{$S'_{21}=S_{20}$}, and for all
\mbox{$i \geq 3$} and \mbox{$j \in \{0,1\}$}, $S'_{ij}=S_{ij}$. We
have that \mbox{$S'_{11}S'_{21}=-S_{10}S_{20}$}, which means that
the answer given by strategy $S'$ on question $x'$ is as
appropriate as the answer given by strategy $S$ on question~$x$.
Moreover,
\begin{eqnarray*}
(S'_{10}+ \imath S'_{11})(S'_{20}+ \imath S'_{21}) & = &
( S_{11}- \imath  S_{10})(-S_{21}+ \imath  S_{20}) \\
&=& -S_{11}S_{21} + \imath S_{11} S_{20}+ \imath  S_{10}S_{21} + S_{10}S_{20} \\
&=& (S_{10}+ \imath S_{11})(S_{20}+ \imath S_{21}) \, .
\end{eqnarray*}
This shows that
\[ \prod_{i=1}^n (S'_{i0}+\imath S'_{i1})
 = \prod_{i=1}^n (S_{i0}+\imath S_{i1}) \, . \]
Since these products are the same, so is their real part, which is
equal to $2^{\left\lfloor n/2 \right\rfloor}$ given that $S$ is optimal.
Therefore, $S'$ is optimal as well.
This establishes that at least as many optimal strategies give the
appropriate answer on $x'$ than on~$x$,
and therefore this number of optimal strategies is the same
for all legitimate questions.
This concludes the proof that the set of all
optimal strategies is balanced, and therefore that
$\mathcal{S}$ is successful with probability~$\optprob$
by virtue of Lemma~\ref{lem:probabilistic}. 
\end{proof}

\section{Imperfect Apparatus}
\label{loophole}

Quantum devices are often unreliable and thus we cannot expect to
witness the perfect results predicted by quantum mechanics in
Theorem~\ref{thm:quant}. However, the \mbox{following} analysis
shows that reasonable imperfections in the apparatus can be
tolerated if we are satisfied with making experiments in which a
quantum-mechanical strategy succeeds with a probability that is
still better than anything classically achievable. Provided care
is taken to make it impossible for the players to ``cheat'' by
communicating after their inputs have been chosen
(see~\cite{BBT04b} for a detailed discussion on this issue), this
would definitely rule out any possible classical (local realistic)
theories of the universe.

First consider the following model of imperfect apparatus.
Assume that the classical bit $y_i$ that is output by each
player $A_i$ corresponds to the predictions of quantum mechanics---should
the apparatus be perfect---with some probability~$p$.
With complementary probability \mbox{$1-p$}, the player
outputs the complement of that bit.
Assume furthermore that the errors are independent between players.
In other words, we model this imperfection as if each player would flip
his (perfect) output bit with probability~\mbox{$1-p$}.
Please note that this assumption of independence does \textit{not} model
imperfections that might occur in the entanglement shared between the
players.

\begin{theorem}\label{BSC}
For any \mbox{$p > \squash{\frac{1}{2}} +
\squash{\frac{\sqrt{2}}{4}} \approx 85\%$} and for any
sufficiently large number $n$ of players, the success probability
of the quantum strategy given in the proof of
\mbox{Theorem}~\ref{thm:quant} for game $G_n$ remains strictly
better than anything classically achievable, provided each player
outputs what is predicted by quantum mechanics with probability at
least~$p$, \mbox{independently} from one another.
\end{theorem}

\begin{proof}
In the $n$-player imperfect quantum strategy, the probability
$p_n$ of winning the game is given by the probability of having
an even number of errors.
\[
 p_n  ~ =  \sum_{i\text{~even}} \!\!
 {\textstyle \binom {n} {i}} \, p^{n-i} (1-p)^{i}
\]
It is easy to prove by mathematical induction that
\[
 p_n \ = \ \frac{1}{2} + \frac{(2p-1)^n}{2} \, .
\]
Let's concentrate for now on the case where $n$ is odd,
in which case \mbox{$\ceil{n/2}=(n+1)/2$}.
By Theorem~\ref{framework:probabilistic},
the success probability of any classical strategy
is upper-bounded by
\[ p'_n =  \frac{1}{2} + \frac{1}{2^{(n+1)/2}} \, .\]
For any fixed $n$, define
\[ e_n = \frac{1}{2} + \frac{(\sqrt2\,)^{1+1/n}}{4} \, . \]
It follows from elementary algebra that
\[
 p > e_n ~\Rightarrow~
 p_n  > p'_n  \, .
\]
In other words, the imperfect quantum strategy on $n$ players
surpasses anything classically achievable provided \mbox{$p >
e_n$}. For example, $e_3 \approx 89.7\%$ and $e_5 \approx 87.9\%$.
Thus we see that even the game with as few as 3 players is
sufficient to \mbox{exhibit} genuine quantum behaviour if the
apparatus is at least 90\% reliable. As~$n$ \mbox{increases}, the
threshold $e_n$ decreases. In~the limit of large~$n$, we have
\[ \lim_{n \rightarrow \infty}  e_n = \frac{1}{2}+ \frac{\sqrt{2}}{4}
\approx 85\% \, .  \] The same limit is obtained for the case when
$n$ is even.
\end{proof}

Another way of modelling an imperfect apparatus is to assume that
it will never give the wrong answer, but that sometimes it fails to
give an answer at all.  This is the type of behaviour that gives rise to
the infamous \emph{detection loophole} in experimental tests
that the world is not classical~\cite{massar} because we say that
the apparatus ``detects'' the correct answer with some probability~$\eta$,
whereas it fails to detect an answer with
complementary probability~\mbox{$1-\eta$}.

To formalize this model, we allow players (classical or quantum) to
answer a special symbol $\bot$ instead of $0$ or~$1$.
We~say that a strategy is \textit{error-free} if, given any
legitimate question, one of two things happens:
\begin{enumerate}
\item at least one player produces $\bot$ as output,
in which case we say that the answer is a~\textit{draw};~or
\item the answer is appropriate for the given question, which can only
happen when none of the players output~$\bot$.
\end{enumerate}
We~say that a player ``provides an output'' whenever that output
is not~$\bot$.
The~larger the probability of obtaining an appropriate answer
for the worst possible question, the better the strategy.
We~are concerned with the smallest possible detection
threshold~$\eta$ that makes a quantum implementation
better than any error-free classical strategy.
But first, we need a Lemma.

\begin{lemma}
Given any classical deterministic error-free strategy for game~$G_n$,
there are at most two legitimate questions on which the players can
provide an appropriate answer.
\end{lemma}

\begin{proof}
Let us dismiss the possibility for some player to output~$\bot$ on
both possible inputs because in that case there would be no
questions at all on which an appropriate answer is obtained.
We~say of a player that he is \textit{interesting} if he never
outputs~$\bot$. For any~$i$, define \mbox{$q_i = \star$} if
player~$i$ is interesting, and otherwise define $q_i$ as the one
input (0~or~1) that results in a non-$\bot$ output for that
player. Consider the string \mbox{$q=q_1 q_2 \cdots q_n$} of
symbols from \mbox{$\{0,1,\star\}$}. We~say that an \mbox{$n$-bit}
string \mbox{$x=x_1 x_2 \cdots x_n$} is \textit{answerable} if
\mbox{$x_i=q_i$} whenever \mbox{$q_i \neq \star$}. The~questions
that give rise to an appropriate answer are precisely those that
are both answerable and legitimate. Let~$\ell$ denote the number
of interesting players. There are $2^\ell$ answerable questions
and exactly half of them are legitimate provided~\mbox{$\ell>0$}.
It~follows that there are $2^{\ell-1}$ legitimate questions on
which the players will provide an appropriate answer.
(If~\mbox{$\ell=0$}, there is only one answerable question, which
may be legitimate or not, and therefore there is at most one
legitimate question on which the players will provide an
appropriate answer.)

Consider any interesting player.
We~say that he is \textit{passive} if his output does not
depend on his input, and that he is \textit{active} otherwise.
Finally, we say that two players are \textit{compatible} either if they
are both active or both passive.
Assume now for a contradiction that \mbox{$\ell \ge 3$}.
Among the $\ell$ interesting players, there must necessarily be
at least two who are compatible; call them Alice and Bob.
Consider any legitimate question that is answerable for which
the input to both Alice and Bob is~0.
(This is always possible by using the degree of freedom provided
by the input to the third interesting player.)
If~we flip the inputs of Alice and Bob, the new question
is still legitimate and still answerable.
The~parity of the answer given by the players on those
two questions is the same because Alice and Bob are compatible.
But~it should \textit{not} be the same because there are two more
1s in the new question.  We~conclude from this contradiction
that~\mbox{$\ell \le 2$}.

The Lemma follows from the fact that there are $2^{\ell-1}$ legitimate
questions on which the players will provide an appropriate answer,
and \mbox{$2^{\ell-1} \le 2$} given that \mbox{$\ell \le 2$}. 
\end{proof}

We now give a simple optimal error-free deterministic strategy
for the game~$G_n$: it~succeeds on exactly two questions.
All~players output~0 on input~0 and $\bot$~on input~1,
except for the first two players.
Player 1 outputs~0 on both inputs and player 2 outputs~0 on input~0
and~1 on input~1.
All~legitimate questions lead to a draw, except
questions $000 \cdots  0$ and $110 \cdots  0$, on which
an appropriate answer is indeed obtained.

\begin{theorem}
For all \mbox{$\eta > \pbfrac{1}{2}$}
and for any sufficiently large number $n$ of
players, the probability that the quantum strategy
given in the proof of Theorem~\ref{thm:quant} for game $G_n$
will produce an appropriate answer
remains strictly better than anything classically achievable
by an error-free strategy,
provided each player outputs what is predicted by quantum
mechanics with probability at least~$\eta$, independently
from one another, and outputs $\bot$ otherwise.
The probabilities are taken according to the uniform distribution
on the set of all legitimate questions.
\end{theorem}

\begin{proof}
There are $2^{n-1}$ legitimate questions and any
classical deterministic error-free strategy
is such that at most two questions give rise to an appropriate answer.
When the questions are asked according to the uniform distribution
on the set of all legitimate questions,
the best a classical deterministic error-free strategy
can do is to provide an appropriate answer with probability
\mbox{$\frac{2}{2^{n-1}}$}. It~is easy to see that
classical \textit{probabilistic} error-free strategies
cannot fare any better.

On~the other hand, if each quantum player from the proof of
Theorem~\ref{thm:quant} outputs the answer predicted by quantum
mechanics with probability~$\eta$ and answers $\bot$ with
complementary probability \mbox{$1-\eta$}, and if these events
are independent, then the probability to obtain an appropriate
answer (none of the players output~$\bot$) is~\mbox{$p_\eta=\eta^n$}.
Elementary algebra suffices to show that \mbox{$p_\eta > \frac{2}{2^{n-1}}$}
precisely when \mbox{$\eta > \frac{\sqrt[n]{4}}{2}$}.
The theorem follows from the fact that
\[ \lim_{n \rightarrow \infty} \frac{\sqrt[n]{4}}{2} = \frac{1}{2} \, . \]
\end{proof}

This result is a significant improvement over~\cite{BM93}, which required
the probability for each quantum player to provide a non-$\bot$ output
to be greater than \mbox{$\frac{1}{\sqrt{2}} \approx 71\%$}
even in the limit of large~$n$.

\section{Conclusions and Open Problems}

We have recast Mermin's $n$-player game into the framework of
pseudo-telepathy, which makes it easier to understand for
non-physicists, and in particular for computer scientists.
An~upper bound was known on the success proportion for any
possible classical deterministic strategy, and therefore
also for the probability of success for any
possible classical probabilistic strategy.
In~this paper, we have proved that these upper bounds are tight.
We~have analysed the issue of when a quantum implementation based on
imperfect or inefficient quantum apparatus remains better than
anything classically achievable.
In~the case of inefficient apparatus, our analysis
provides a significant \mbox{improvement} on what was previously known.

A lot is known about pseudo-telepathy~\cite{BBT04b} but many questions
remain open.  The game studied in this article has been
generalized to larger inputs~\cite{zuk,BHMR03} and
larger outputs~\cite{Bo04}.
It~would be interesting to have tight bounds for those more general games.
Also, it would be interesting to know how to construct the
pseudo-telepathy game that minimizes classical success probability
when the dimension of the entangled quantum state is fixed.
In~all the pseudo-telepathy games known so far, it is sufficient
for the quantum players to perform a projective von Neumann measurement.
Could there be a \textit{better} pseudo-telepathy game (in~the sense
of making it harder on classical players) that would make inherent use
of generalized measurements~(POVM)?

We have modelled imperfect apparatus in two different ways: when they
produce incorrect outcomes and when they don't produce outcomes at all.
It~would be natural to combine those two models into a more realistic one,
in which each player receives an outcome with probability~$\eta$,
but that outcome is only correct with probability~$p$.
Finally, we should model other types of errors in the quantum
process, such as imperfections in the prior entanglement shared
among the players.

\section*{Acknowledgements}
We are most grateful to Serge Massar for pointing out Mermin's
paper~\cite{Me90} after we thought we had invented this game~\cite{WADS}!
We are also grateful to George Savvides for helping us simplify
some of our proofs.

\bibliographystyle{splncs}

\end{document}